\documentclass[russian,english]{article}
\usepackage[T2A,T2A,T2A,T1]{fontenc}
\usepackage[latin9]{inputenc}
\usepackage{amsmath}
\usepackage{amssymb}
\usepackage{esint}

\makeatletter

\DeclareRobustCommand{\cyrtext}{%
  \fontencoding{T2A}\selectfont\def\encodingdefault{T2A}}
\DeclareRobustCommand{\textcyr}[1]{\leavevmode{\cyrtext #1}}
\AtBeginDocument{\DeclareFontEncoding{T2A}{}{}}

\newcommand{\lyxmathsym}[1]{\ifmmode\begingroup\def\b@ld{bold}
  \text{\ifx\math@version\b@ld\bfseries\fi#1}\endgroup\else#1\fi}

\@ifundefined{date}{}{\date{}}
\AtBeginDocument{\DeclareFontEncoding{T2A}{}{}}\AtBeginDocument{\DeclareFontEncoding{T2A}{}{}}\usepackage{indentfirst}\usepackage{cmap}

\newtheorem{theorem}{Theorem}
\newtheorem{corollary}{Corollary}
\newtheorem{lemma}{Lemma}

\newtheorem{remark}{Remark}

\usepackage{babel}

\usepackage{babel}

\usepackage{babel}

\makeatother

\usepackage{babel}
\begin{document}

\title{Phase diagram for one-way traffic flow with local control }

\author{Lykov A. A., Malyshev V. A., Melikian M.V. \thanks{Lomonosov Moscow State University, Faculty of Mechanics and Mathematics, Vorobyevy Gory 1, Moscow, 119991, Russia}}
\maketitle
\begin{abstract}
We consider one-way road deterministic traffic model with $N$ particles.
The simplest local control protocol, which reminds physical interaction)\d{ }with
three parameters is considered. We study the stable and unstable domains
of the phase diagram \.{u}niformly in $N$.
\end{abstract}

\section{Introduction}

Theoretical modelling and computer simulation of transportation systems
is a very popular field, see very impressive review \cite{Helbing}.
There are two main directions in this research - macro and micro models.
Macro approach does not distinguish individual transportation units
and uses analogy with the fluid flow in hydrodynamics, see \cite{Prigogine}.
Stochastic micro models are most popular and use almost all types
of stochastic processes: mean field, queueing type and local interaction
models. We consider here completely deterministic transportation flows.
Although not as popular as stochastic traffic, there is also a big
activity in this field, see \cite{feintuch,Hui_Berg,jovanovic_Bamich,Melzer_Kuo,Swaroop}.
In these papers interesting results are obtained for sufficiently
general protocols\foreignlanguage{russian}{. A popular topics in many
papers is to consider optimization of some functional of the chain
of cars. For example, in \cite{jovanovic_Bamich,Melzer_Kuo} and in
references therein, the following model is considered 
\[
\ddot{z}_{n}+\alpha\dot{z}_{n}=u_{n},\quad n=1,\ldots,M,
\]
where $\alpha>0$, $z_{n}=z_{n}(t)$ is the coordinate of the $n$-th
car on the real axis, $u_{1}(t),u_{2}(t),\ldots,u_{M}(t)$ --- control
functions, which should be chosen to minimize the functional 
\[
J=\frac{1}{2}\int_{0}^{\infty}\sum_{n=1}^{M+1}a(z_{n-1}(t)-z_{n}(t)-d)^{2}+b(\dot{z}_{n}(t)-v)^{2}+c(u_{n}(t)-\alpha v)^{2}dt,\quad u_{M+1}=\alpha v.
\]
where $a,b,c$ - some-positive constants, $v,d$ are correspondingly
safe velocity and distance between cars. Moreover, it is assumed that
\[
z_{0}(t)=vt,\ z_{M+1}(t)=vt-(M+1)d,
\]
It is clear however, that any model should satisfy some natural necessary
conditions. For example, besides minimizing such functionals\c{ }
the inequality should hold 
\[
I=\inf_{t,n}(z_{n-1}(t)-z_{n}(t))>0
\]
that is necessary for safety.}

Here we follow another strategy: for simplest possible protocols we
try to get results as concrete as possible. Namely, we consider the
one-way road traffic model organized as follows.

At any time $t\geq0$ there is finite or infinite number of point
particles (may be called also cars, units etc.) with coordinates $z_{k}(t)$
on the real axis, enumerated as follows 
\begin{equation}
...<z_{n}(t)<...<z_{1}(t)<z_{0}(t)\label{sequence}
\end{equation}
We assume that the rightmost car (the leader) moves ``as it wants'',
that is the trajectory $z_{0}(t)$ is often assumed to have non\={n}egative
velocity.

Our problem is to find the simplest possible local protocol (control
algorithm) which would guarantee both safety (no collisions), stable
(or even maximal) density of the flow or maximal current. Otherwise
speaking, we try to find control mechanism which guarantees that the
distance between any pair of neighbouring cars is close (on all time
interval $(0,\infty)$) to some (given a priori) fixed number, that
defines the density of the flow.

More exactly, denoting $r_{k}(t)=z_{k-1}(t)-z_{k}(t)$, and

\[
I=\inf_{k\geqslant1}\inf_{t\geqslant0}r_{k}(t),\,\,\,S=\sup_{k\geqslant1}\sup_{t\geqslant0}r_{k}(t),
\]
we try to get the bounds - lower positive bound on $I$ and upper
bound on $S$ - as close as possible.

Locality (of the control) means that the ``driver'' of the $k$-th
car, at any time $t$, knows only its own velocity $v_{k}(t)$ and
the distance $r_{k}(t)$ from the previous car. Thus, for any $k\geq1$
the trajectory $z_{k}(t)$, being deterministic, is uniquely defined
by the trajectory $z_{k-1}(t)$ of the previous particle.

Using physical terminology one could say that if, for example, $r_{k}(t)$
becomes larger than $d$, then some virtual force $F_{k}$ increases
acceleration of the particle $k$, and vice-versa. Thus the control
mechanism is of the physical nature, like forces between molecules
in crystals but our ''forces'' are not symmetric. Thus our system
is not a hamiltonian system. Nevertheless, our results resemble the
dynamical phase transition in the model of the molecular chain rapture
under the action of external force, see \cite{Mal_Muz}. However here
we do not need the double scaling limit used in \cite{Mal_Muz}.

We will see however that for the stability, besides $F_{k}$, also
friction force $-\alpha v_{k}(t)$, restraining the growth of the
velocity $v_{k}(t)$, is necessary, where the constant $\alpha>0$
should be chosen appropriately. Taking $F_{k}$ to be simplest possible
\begin{equation}
F_{k}(t)=\omega^{2}(z_{k-1}(t)-z_{k}(t)-d)\label{linear_force}
\end{equation}
we get that the trajectories are uniquely defined by the system of
equations for $k\geq1$ 
\begin{equation}
\frac{d^{2}z_{k}}{dt^{2}}=F_{k}(t)-\alpha\frac{dz_{k}}{dt}=\omega^{2}(z_{k-1}(t)-z_{k}(t)-d)-\alpha\frac{dz_{k}}{dt}\label{main_equation}
\end{equation}

Stability depends not only on the parameters $\alpha,\omega,d$ but
also on the initial conditions and on the movement of the leader (on
its velocity and acceleration). This is easy to understand for the
case of $N+1$ particles. For example, for $N=1$, where the calculations
are comp\={l}etely trivial, assume also the simplest leader movement
\begin{equation}
z_{0}(t)=vt,t\geq0\label{z_0_vt}
\end{equation}
Then, if initial condition for the second particle are 
\[
z_{1}(0)=-a=-(d+\frac{\alpha}{\omega^{2}}v),\dot{z}_{1}(0)=v,
\]
then $z_{1}(t)=-a+vt$ for any $d,\alpha,\omega$. However, if we
change only the initial velocity $\dot{z}_{1}(0)=w$ to some $w>0$,
then for any $\alpha,\omega$ there exists $w_{1}=w_{1}(\alpha,\omega,d)$
such that for any $w\geq w_{1}$ collision occurs.

For $N=2,3,...$ the situation becomes more and more complicated,
and its study has no much sense. That is why we study, in the space
of two parameters $\alpha,\omega$ (for fixed $d$), stability conditions,
which are uniform in $N$ and in large class of reasonable initial
conditions and reasonable movement of the leader.

Natural (reasonable) initial conditions are as follows: at time $0$
it should be 
\[
0<\inf_{k\geqslant1}r_{k}(0)\leq\sup_{k\geqslant1}r_{k}(0)<\infty
\]
As for the leader movement, it is sometimes sufficient to assume that
the function $z_{0}(t)$ were continuous, but in other cases it is
assumed to twice differentiable and has the following bounds on the
velocity and acceleration of the leader: 
\begin{equation}
\sup_{t\geqslant0}|\dot{z}_{0}(t)|=v_{\mathrm{max}},\quad\sup_{t\geqslant0}|\ddot{z}_{0}(t)|=a_{\mathrm{max}},\quad\label{leader_bounds}
\end{equation}

It appears that under these conditions there are 3 sectors in the
quarter-plane $R_{+}^{2}=\{(\alpha,\omega)\}$: 1) $\alpha>2\omega$,
where we can prove stability, 2) $\alpha<\sqrt{2}\omega$, where we
can prove instability, and the sector 3) $\sqrt{2}\omega\leq\alpha\leq2\omega$,
where we can prove stability only for more restricted classes of initial
conditions and of the leader motion.

\section{Results}

\subsection{Stability}

Here we consider the region $\alpha>2\omega$, and this is always
assumed in this section.

\paragraph{Movement close to stationary}

For any given $d>0$ there are special initial conditions (which can
be called equilibrium configuration) when the force acting on any
particle is zero: 
\begin{equation}
z_{k}(0)=-ka,\ \dot{z}_{k}(0)=v,\ k=0,1,2,\ldots,\label{a_k}
\end{equation}
with 
\begin{equation}
a=a(\alpha,\omega,d,v)=d+\frac{\alpha}{\omega^{2}}v\label{def_a}
\end{equation}
If the leader moves as (\ref{z_0_vt}), then for any $k>0$ also 
\[
z_{k}(t)=z_{k}(0)+vt
\]
Such movement we call stationary. Now consider a perturbation of this
situation

\begin{theorem} \label{Th_stability_1} For given $d$ and $v$ assume
\begin{equation}
|z_{k}(0)+ka|\leqslant\theta a,\quad|\dot{z}_{k}(0)-v|\leqslant\beta v,\label{pr_stab}
\end{equation}
\begin{equation}
\sup_{t\geqslant0}|z_{0}(t)-vt|\leqslant\delta a\label{leader_move}
\end{equation}
for $0\leq\delta<\frac{1}{2}$ and $\theta,\beta\geqslant0$ such
that 
\begin{equation}
\epsilon=2\max\{\delta,\frac{\theta+\frac{2v\beta}{\alpha a}}{\sqrt{1-\left(\frac{2\omega}{\alpha}\right)^{2}}}\}<1\label{epsilon}
\end{equation}
Then 
\begin{equation}
(1-\epsilon)a\leqslant I\leqslant S\leqslant(1+\epsilon)a\label{I_S_bounds_3}
\end{equation}

\end{theorem}

Now we want to show that in some cases the sufficient conditions of
this theorem are are not very far from necessary. For example, from
the theorem follows that the condition 
\begin{equation}
\theta<\frac{1}{2}\sqrt{1-\left(\frac{2\omega}{\alpha}\right)^{2}},\delta=\beta=0\label{etacond}
\end{equation}
is sufficient for $I>0$. This means that the deviations of the particle
from the equilibrium (that is the distances between particles are
equal to $a$) do not exceed $\frac{a}{2}$.

\begin{remark}\label{remark_finite_dim}

The movement of any first $N$ particles does not depend on other
particles. Thus for any $N$ the system of equations (\ref{main_equation})
for $k=1,...,N,$ is finite-dimensional, linear and, for any $\alpha,\omega>0$,
the spectrum of this linear operator is inside the left half-plane.
One could look at it formally, neglecting possible collisions between
particles. In this case it is asymptotically stationary, that is,
as $t\to\infty$, its solution converges to stationary for any initial
conditions and if $z_{0}(t)=vt$. However we do not know whether collisions
occur before it becomes stationary. In the following theorem we get
such conditions.

Because of this one could think that the spectrum does not play big
role in the stability problems. This is not quite true if we consider
the problems uniform in $N$, or the corresponding infinite-dimentional
operator, see below. 

Note also that for the random movement of the leader the same results
hold. More exactly, assume that $z_{0}(t)$ is a stationary process
with smooth trajectories, satisfying bounds (\ref{leader_bounds})
with probability one. Then the mean velocity of any particle converges
to the mean velocity of the leader.

\end{remark}

\begin{theorem} \label{Th_stability_2}

1) Assume (\ref{a_k}) and (\ref{leader_move}). Then for all $k=1,2,\ldots$
we have:

$\sup_{t\geqslant0}|\dot{z}_{k}(t)-v|\leqslant\sup_{t\geqslant0}|\dot{z}_{0}(t)-v|.$
If the righthand side is sufficiently small then $v_{k}(t)=\dot{z}_{k}(0)>0$
for any $k,t$.

2) Assume (\ref{z_0_vt}), (\ref{pr_stab}) and that for some parameters
$\theta,\beta>0$ 
\begin{equation}
\zeta=2\frac{\theta+\frac{2v\beta}{\alpha a}}{\sqrt{1-\left(\frac{2\omega}{\alpha}\right)^{2}}}<1\label{dzeta_less_1}
\end{equation}
Then for all $k$ 
\[
\lim_{t\rightarrow\infty}(z_{k}(t)-(vt-ka))=0
\]
where $a$ is in (\ref{def_a}) and 
\[
(1-\zeta)a\leqslant I\leqslant S\leqslant(1+\zeta)a.
\]
\end{theorem}

\paragraph{Non-stationary initial conditions}

In Theorems (\ref{Th_stability_1}) and (\ref{Th_stability_2}) we
considered initial coordinates of the particles close to the fixed
lattice points $-ka$. Here we consider more natural initial conditions
with restrictions only on the distances between particles. Denote
\begin{equation}
d^{*}=\frac{a_{\mathrm{max}}+\alpha v_{\mathrm{max}}}{\omega^{2}}\label{d_star}
\end{equation}

\begin{theorem} \label{Th_stability_3} Let the initial conditions
be 
\begin{equation}
(1-\theta)d\leqslant r_{k}(0)\leqslant(1+\theta)d,\quad|\dot{z}_{k-1}(0)-\dot{z}_{k}(0)|\leqslant\beta,\quad k=1,2,\ldots.\label{init_cond_b}
\end{equation}
for some $\beta\geqslant0,0\leqslant\theta<1$. Assume moreover that\foreignlanguage{russian}{
\begin{equation}
\eta=\max(\frac{d^{*}}{d},\frac{\theta+\frac{2\beta}{\alpha d}}{\sqrt{1-(\frac{2\omega}{\alpha})^{2}}})<1\label{eta_less_1}
\end{equation}
}Then we have the following stability bounds 
\begin{equation}
(1-\eta)d\leqslant I\leqslant S\leqslant(1+\eta)d\label{I_S_bounds-1}
\end{equation}

\end{theorem}

\begin{remark}\label{remark_th3}

If $\beta=\theta=0$, one can prove the same result for the case $\alpha=2\omega$.

\end{remark}

\begin{corollary} \label{Corrollary_th3} Assume that 
\[
r_{k}(0)=z_{k-1}(0)-z_{k}(0)=d,\quad\dot{z}_{k}(0)=\dot{z}_{0}(0),\quad k=1,2,\ldots.
\]
If moreover 
\[
d^{*}<d
\]
then 
\begin{equation}
d-d^{*}\leqslant I\leqslant S\leqslant d+d^{*}\label{I_S_bounds_2}
\end{equation}

\end{corollary}

\begin{remark}\label{remark_general_stab} It is natural to consider
the stability problem for more general initial conditions. Assume
we know the initial conditions and the parameters $v_{\mathrm{max}},a_{\mathrm{max}}$
of the leader. We want to know whether the parameters $\alpha,\omega,d$
such that $0<I\leqslant S<\infty$, exist. Namely, we will prove that
for any $z_{0}(t)$, satisfying (\ref{leader_bounds}) and the initial
conditions such that: 
\[
\min_{k\geqslant1}(z_{k-1}(0)-z_{k}(0))=A>0,\quad\max_{k\geqslant1}(z_{k-1}(0)-z_{k}(0))=B<\infty,\quad\max_{k\geqslant1}|\dot{z}_{k-1}(0)-\dot{z}_{k}(0)|=C<\infty,
\]
the parameters $\alpha,\omega,d$, such that $0<I\leqslant S<\infty$,
always exist.

\end{remark}

Theorem \ref{Th_stability_1}, \ref{Th_stability_3} and Corollary
\ref{Corrollary_th3}

1) guarantee that there are no collisions between particles (by the
lower bound),

2) provide lower bound for the density of the flow (by the upper bound),
that is the mean distance between particles remains bounded.

3) do not guarantee that the velocities remain positive. Such conditions
were obtained above in Theorem \ref{Th_stability_2}.

\paragraph{Flow density}

Let $n(t,I)$ be the number of units on the interval $I\subset R$
at time $t$. Instead of the density ($|I|$ is the length of $I$)
\[
\rho(t)=\liminf_{|I|\to\infty}\frac{n(t,I)}{|I|}
\]
it is more convenient to consider the inverse density, or the mean
length of the chain of cars $0,1,...,N$ 
\[
L_{N}(t)=\frac{z_{0}(t)-z_{N}(t)}{N}.
\]

\begin{theorem} \label{th_dens}

For any $\alpha>0,\omega>0$ assume that the initial conditions (\ref{init_cond_b})
are such that the following finite limits exist 
\[
\lim_{N\rightarrow\infty}L_{N}(0)=L(0),\quad\lim_{N\rightarrow\infty}\dot{L}_{N}(0)=\dot{L}(0),
\]
Then for any $t$ there exists 
\[
\lim_{N\rightarrow\infty}L_{N}(t)=L(t),
\]
and moreover 
\[
L(t)=L(0)+\frac{1}{\alpha}(1-e^{-\alpha t})\dot{L}(0)
\]
\end{theorem}

Note that if moreover $\frac{dz_{k}(0)}{dt}$ are uniformly bounded
then $\frac{dL}{dt}(0)=0$, that is the mean length does not change
with time.

Due to convergence to stationary movement, the flow current converges,
at any point, to the current of the stationary flow 
\[
va^{-1}=\frac{\omega^{2}v}{\omega^{2}d+\alpha v}
\]

\paragraph{Restricted stability}

Here we consider the region $\sqrt{2}\omega\leq\alpha\leq2\omega$,
where we can prove stability only for asymptotically homogeneous initial
conditions.

\selectlanguage{russian}%
\begin{theorem}\label{Th_stability_4}\par Let $\sqrt{2}\omega\leqslant\alpha\leq2\omega$
and let $z_{0}(t)$ be such that 
\[
\omega\int_{0}^{\infty}|z_{0}(t)-vt|dt=\sigma a<\infty
\]
for some $\sigma\geqslant0$. Assume also that the initial conditions
are ``summable'', that is 
\[
\sum_{k=1}^{\infty}|z_{k}(0)+ka|\leqslant\theta a,\quad\sum_{k=1}^{\infty}|\dot{z}_{k}(0)-v|\leqslant\beta v
\]
for some $\theta,\beta\geqslant0$. Then 
\[
I\geqslant(1-2\eta)a,\quad S\leqslant(1+2\eta)a,
\]
where 
\[
\eta=2\left(\theta+\frac{\beta v}{a\omega}+\sigma\right)
\]
\end{theorem}

It follows from this theorem that the upper bound $S<\infty$ holds
for all parameters, but the lower (safety) bound $I>0$ holds if 
\[
\theta+\frac{\beta v}{a\omega}+\sigma<\frac{1}{4}.
\]

\begin{theorem}\label{Th_stability_5}\par Assume again that $\sqrt{2}\omega\leqslant\alpha\leq2\omega$,
the initial conditions satisfy 
\[
\sum_{k=1}^{\infty}|r_{k}(0)-d|\leqslant\theta d,\quad\sum_{k=1}^{\infty}|\dot{z}_{k-1}(0)-\dot{z}_{k}(0)|\leqslant\beta,\quad k=1,2,\ldots.
\]
for some $\beta\geqslant0,0\leqslant\theta\leqslant1$, and the leader
moves as 
\[
\frac{1}{\omega}\int_{0}^{+\infty}|\ddot{z}(t)+\alpha\dot{z}_{0}(t)|dt=\sigma d<\infty
\]
for some $\sigma\geqslant0$. Then 
\[
I>d(1-\eta),\quad S<d(1+\eta),
\]
where 
\[
\eta=2\left(\theta+\frac{\beta}{\omega d}+\sigma\right),\quad x=\frac{\alpha}{\omega}.
\]
\end{theorem}

It follows that the safety condition $I>0$ holds if 
\[
\theta+\frac{\beta}{\omega d}+\sigma<\frac{1}{2}.
\]

\selectlanguage{english}%

\subsection{Instability}

Here we will prove instability for the region $\alpha<\sqrt{2}\omega$.
The first reason for the instability is the absence of dissipation,
that is if $\alpha=0$. The following result shows this even for the
most favorable initial conditions.

\begin{theorem}\label{th_instab_1}

Assume that $\alpha=0$ and 
\begin{equation}
z_{k}(0)=-ka,\,\frac{dz_{k}}{dt}(0)=v,\,k\geq0\label{initial_simplest}
\end{equation}
\[
z_{0}(t)=tv+\sin\omega_{0}t,\ v>0,\ \omega_{0}\ne0
\]
Then for any $k\geqslant2$ we have due to resonance 
\[
\inf_{t\geqslant0}r_{k}(t)=-\infty
\]
\end{theorem}

Now we show that, even under the smallest perturbation of the initial
conditions (\ref{initial_simplest}) and even simpler leader trajectory,
we get much more general instability condition.

\begin{theorem} \label{th_instab_2} Assume $\alpha<2\omega$, $z_{0}(t)=vt$
and initial conditions such that: 
\begin{equation}
z_{k}(0)=-ka,\ \dot{z}_{k}(0)=\begin{cases}
v, & k>1\\
v+\epsilon, & k=1
\end{cases},\label{eq:asymp}
\end{equation}
where $\epsilon$ is some real number. Then for any $\mu>\frac{1}{\tau},\ \tau=\sqrt{\omega^{2}-\frac{\alpha^{2}}{4}}$
the following asymptotic formula takes place: 
\[
z_{k+1}(t)-(vt-(k+1)a)\sim\frac{c}{\sqrt{k}}e^{kf(\mu)}\sin(\Omega(\mu)k+\phi_{0}(\mu)),\ \mbox{when}\ t=\mu k,\ k\rightarrow\infty
\]
where 
\[
f(\mu)=-\frac{\alpha\mu}{2}+1-\ln\left(\frac{2\tau}{\mu\omega^{2}}\right),
\]
\[
\phi_{0}(\mu)=\arctan(\nu),\nu=\sqrt{\mu^{2}\tau^{2}-1},\Omega(\mu)=\nu-\phi_{0}(\mu)=\nu-\arctan(\nu),c=\epsilon\sqrt{\frac{2\tau}{\pi\nu\mu}}
\]
\end{theorem}

\begin{corollary} \label{-Corollary_2} Assume $\alpha<\sqrt{2}\omega$
and $z_{0}(t)=vt$. Assume the initial conditions (\ref{eq:asymp}).
Then

\[
I=-\infty,\quad S=\infty.
\]
\end{corollary}

While proving the theorem we will see that corollary (\ref{-Corollary_2})
holds even for more general initial conditions: 
\[
\max_{k}|z_{k}(0)+ka|\leqslant\epsilon_{1},\quad\max_{k}|\dot{z_{k}}(0)-v|\leqslant\epsilon_{2}.
\]
with some nonnegative $\epsilon_{1},\epsilon_{2}\geqslant0$.

\section{Stability: proofs}

\subsection{Theorem \ref{Th_stability_3}, Corollary \ref{Corrollary_th3}, Remark
\ref{remark_finite_dim}}

\selectlanguage{russian}%
It is very convenient to use new variables 
\begin{equation}
x_{0}(t)=\frac{\ddot{z}_{0}(t)+\alpha\dot{z}_{0}(t)}{\omega^{2}},\,\,\,x_{k}(t)=z_{k-1}(t)-z_{k}(t)-d=r_{k}(t)-d,\quad k=1,2,\ldots\label{def_x_k}
\end{equation}
Using 
\[
\ddot{z}_{k}=\omega^{2}x_{k}-\alpha\dot{z}_{k}=\omega^{2}x_{k}-\alpha(\dot{z}_{k-1}-\dot{x}_{k}),\ddot{z}_{k}=\ddot{z}_{k-1}-\ddot{x}_{k}
\]
we get the main equations (\ref{main_equation}) in terms of these
variables: 
\begin{equation}
\ddot{x}_{k}+\alpha\dot{x}_{k}+\omega^{2}x_{k}=\ddot{z}_{k-1}+\alpha\dot{z}_{k-1}=\omega^{2}x_{k-1},\quad x_{0}=\frac{\ddot{z}_{0}+\alpha\dot{z}_{0}}{\omega^{2}}.\label{x_main}
\end{equation}

\begin{lemma}\label{mainLemma} Assume that 
\[
\sup_{t\geqslant0}|x_{0}(t)|=Q<\infty.
\]
Let also the initial conditions be such that 
\[
\sup_{k\geqslant1}|x_{k}(0)|=A<\infty,\quad\sup_{k\geqslant1}|\dot{x}_{k}(0)|=C<\infty.
\]
If $\alpha>2\omega$ then 
\[
\sup_{k,t}|x_{k}(t)|\leqslant Q'=\max\{Q,\frac{\alpha A+2C}{\sqrt{\alpha^{2}-4\omega^{2}}}\}.
\]
\par \end{lemma}

Proof. The solution of the linear equation (\ref{x_main}) for any
fixed $k$ is of course well-known. Namely, if the roots 
\[
\lambda_{\pm}=-\frac{\alpha}{2}\pm\sqrt{\frac{\alpha^{2}}{4}-\omega^{2}}
\]
of the characteristic equation 
\begin{equation}
G(\lambda)=\lambda^{2}+\alpha\lambda+\omega^{2}=0\label{G_x}
\end{equation}
are different, then the solution can be written as 
\[
x_{k}(t)=x_{k,+}(t)+x_{k,-}(t)
\]
where 
\[
x_{k,\pm}(t)=C_{\lyxmathsym{\textcyr{\char234}},\pm}e^{\lambda_{\pm}t}+\omega^{2}\frac{e^{\lambda_{\pm}t}}{Q'(\lambda_{\pm})}\int_{0}^{t}e^{-\lambda_{\pm}t_{1}}x_{k-1}(t_{1})dt_{1}=
\]
\[
=C_{\lyxmathsym{\textcyr{\char234}},\pm}e^{\lambda_{\pm}t}+\omega^{2}\frac{e^{\lambda_{\pm}t}}{2\lambda_{\pm}+\alpha}\int_{0}^{t}e^{-\lambda_{\pm}t_{1}}x_{k-1}(t_{1})dt_{1}
\]
Using the initial conditions one finds 
\[
C_{k,\pm}=\frac{1}{2\gamma}((\pm\frac{\alpha}{2}+\gamma)a_{k}\pm b_{k}),\,\,\,\gamma=\sqrt{\frac{\alpha^{2}}{4}-\omega^{2}},
\]
where $x_{k}(0)=a_{k},\,k\in\mathbb{N}$, $\dot{x}_{k}(0)=b_{k},k\in\mathbb{N}$.
As $\lambda_{+}>\lambda_{-}$ we get: 
\[
|x_{k}(t)|\leq|C_{\lyxmathsym{\textcyr{\char234}},+}|\exp(\lambda_{+}t)+|C_{k,-}|\exp(\lambda_{-}t)+\sup_{s\geq0}|x_{k-1}(s)|\frac{\omega^{2}}{2\gamma}\int_{0}^{t}|\exp(\lambda_{+}(t-t_{1}))-\exp(\lambda_{-}(t-t_{1}))|dt_{1}
\]
Let us prove that for $y_{k}=\sup_{s\geq0}|x_{k}(s)|$ the following
inequalities hold 
\begin{equation}
y_{k}\leq\max\{y_{k-1},\,\frac{\alpha A+2C}{2\gamma}\}.\label{reqq_y}
\end{equation}
Putting $t-t_{1}=s$ and using $\omega^{2}=\lambda_{+}\lambda_{-}$
and\foreignlanguage{english}{ 
\[
|C_{k,\pm}|\leq\frac{(\frac{\alpha}{2}\pm\gamma)A+C}{2\gamma}
\]
}we get: 
\[
|x_{k}(t)|\leq|C_{\lyxmathsym{\textcyr{\char234}},+}|\exp(\lambda_{+}t)+|C_{k,-}|\exp(\lambda_{-}t)+\sup_{s\geq0}|x_{k-1}(s)|\frac{\omega^{2}}{2\gamma}\int_{0}^{t}(\exp(\lambda_{+}s)-\exp(\lambda_{-}s))ds\leq
\]
\begin{equation}
\leq\frac{1}{2\gamma}((\frac{\alpha}{2}+\gamma)A+C+y_{k-1}\lambda_{-})\exp(\lambda_{+}t)+\frac{1}{2\gamma}((\frac{\alpha}{2}-\gamma)A+C-y_{k-1}\lambda_{+})\exp(\lambda_{-}t)+y_{k-1}\label{x_recurrence}
\end{equation}
To finish the proof we need the following simple lemma.

\begin{lemma}\label{f_t_max}Let us consider the function 
\[
f(t)=a\exp(\lambda_{+}t)+b\exp(\lambda_{-}t)+c
\]
for some constants $b,\,c>0$, $a\in\mathbb{R}$, $\lambda_{-}<\lambda_{+}<0$.
For every $t\geq0$ the following statement is true: 
\[
|f(t)|\leq\max\{c,\,a+b+c\}.
\]
\par \end{lemma}

Proof. There are two cases:

1. $a>0$, then it is obvious that $\sup_{s\geq0}|f(s)|=f(0)=a+b+c$.

2. $a<0$, in this case the set of such $t$, that the derivative
of the function is positive $0$, is:

\[
t>\frac{1}{\lambda_{-}-\lambda_{+}}\ln(-\frac{a}{b}\frac{\lambda_{+}}{\lambda_{-}})
\]
The point $t_{0}=\frac{1}{\lambda_{-}-\lambda_{+}}\ln(-\frac{a}{b}\frac{\lambda_{+}}{\lambda_{-}})$
is the minimum point. If $t_{0}<0$, then $\sup_{s\geq0}|f(s)|$ tops
in the point $+\infty$ and is equal to $c$. If $t_{0}>0$, then
$\sup_{s\geq0}|f(s)|$ tops in the point $0$ or in the point $+\infty$.

Having applied this lemma to (\ref{x_recurrence}) we get (\ref{reqq_y})
and 
\[
|x_{k}(t)|\leq\max\{Q,\,B\}=Q^{'}
\]
where $B=\frac{\alpha A+2C}{2\gamma}$. Lemma \ref{mainLemma} is
thus proved.

To prove theorem \foreignlanguage{english}{\ref{Th_stability_3} it
is sufficient to put $A=\theta d,\,C=\beta,\,Q=d^{*}$. To prove corollary
\ref{Corrollary_th3} we can put $\theta=\beta=0$.}

\selectlanguage{english}%

\paragraph{Remark}

\selectlanguage{russian}%
We use here theorem \ref{Th_stability_3} and the notation therein.
From 
\begin{equation}
(1-\theta)d=A,\quad(1+\theta)d=B,\quad\beta=C,\label{thcondeq}
\end{equation}
we find first the parameters $d,\theta,$ 
\[
d=\frac{A+B}{2},\quad\theta=\frac{B-A}{2d}=\frac{B-A}{A+B},.
\]
Then we find $\alpha,\omega$ such that $\alpha>2\omega$ and $\eta<1$.
Namely, we choose $\alpha$ sufficiently large and take any $\omega$
so that 
\[
\frac{1}{\sqrt{d}}\sqrt{a_{\mathrm{max}}+\alpha v_{\mathrm{max}}}<\omega<\frac{\alpha}{2}
\]
Then the inequality (\ref{eta_less_1}) evidently holds.

\selectlanguage{english}%

\subsection{Theorem \ref{Th_stability_1}}

\selectlanguage{russian}%
Note that the functions $q_{k}(t),k=1,2...$, defined by 
\begin{equation}
z_{k}(t)=vt-ka+q_{k}(t),\label{solDec}
\end{equation}
satisfy the system of equations (\ref{x_main}) with the initial conditions:
\[
q_{0}(t)=z_{0}(t)-vt,\,\,\,q_{k}(0)=z_{k}(0)+ka,\quad\dot{q}_{k}(0)=\dot{z}_{k}(0)-v,\,\,\,k=1,2...
\]
Then 
\[
\sup_{k}|q_{k}(0)|\leqslant\theta a,\quad\sup_{k}|\dot{q}_{k}(0)|\leqslant\beta v,\quad\sup_{t}|q_{0}(t)|=\delta a.
\]
Using lemma \ref{mainLemma} one has the lower bound 
\[
r_{k}(t)=z_{k-1}(t)-z_{k}(t)=a+q_{k-1}-q_{k}(t)\geqslant a-2\sup_{k,t}|q_{k}(t)|=a(1-\epsilon).
\]
The upper bound for $r_{k}(t)$ can be obtained similarly.

\selectlanguage{english}%

\subsection{Theorem \ref{Th_stability_2}}

\selectlanguage{russian}%
Differentiating the system (\ref{x_main}) we get the equations for
velocities $v_{k}(t)=\dot{z}_{k}(t).$ 
\begin{equation}
\ddot{v}_{k}=\omega^{2}(v_{k-1}-v_{k})-\alpha\dot{v}_{k},\label{velEq}
\end{equation}
with initial conditions: 
\[
v_{k}(0)=v,\ \dot{v}_{k}(0)=0.
\]
So for every $k=0,1,2,\ldots$: 
\[
v_{k}(t)=v+u_{k}(t),
\]
where $u_{k}$ satisfy equations (\ref{velEq}) with zero initial
conditions $u_{0}(t)=v_{0}(t)-v,\,\,u_{k}(0)=\dot{u}_{k}(0)=0,\,\,k\geq1$.
So from lemma ?\ref{mainLemma} for $u_{k}$ we get that for \t{a}ny
$k=1,2,\ldots$ and $t\geqslant0$ 
\[
|u_{k}(t)|\leqslant\sup_{t\geqslant0}|u_{0}(t)|=\sup_{t\geqslant0}|\dot{z}_{0}(t)-v|
\]
This proves assertion 1) of the Theorem.

Let us prove now assertion 2). The functions $q_{k}(t)$ defined in
(\ref{solDec}) satisfy the system (\ref{x_main}) with initial conditions:
\[
q_{k}(0)=z_{k}(0)+ka,\quad\dot{q}_{k}(0)=\dot{z}_{k}(0)-v.
\]
Moreover $q_{0}(t)=z_{0}(t)-vt=0$ for all $t\geqslant0$. By induction
we get: 
\[
\left(\frac{d^{2}}{dt^{2}}+\alpha\frac{d}{dt}+\omega^{2}\right)^{k}q_{k}=0.
\]
which gives 
\[
q_{k}(t)=e^{\lambda_{1}t}P_{k}(t)+e^{\lambda_{2}t}Q_{k}(t),
\]
where $P_{k},Q_{k}$ are the polynomials of degree $k-1$ and $\lambda_{1},\lambda_{2}$
are the roots of the equation (\ref{G_x}). Note that, for any $\omega,\alpha>0$,
the real parts of $\lambda_{1},\lambda_{2}$ are negative.

The last statement follows if we put $\delta=0$ in (\ref{epsilon}),
which gives (\ref{dzeta_less_1}).

\subsection{Theorems \ref{Th_stability_4} and \ref{Th_stability_5}}

Before proving them, we need new notation. \foreignlanguage{english}{Introduce
the Banach space 
\begin{equation}
X=\{\psi=(q,p),\ q=\{q_{k}\}_{k=1,2\ldots},\ p=\{p_{k}\}_{k=1,2\ldots}:\ ||q||=\sup_{k}|q_{k}|<\infty,\ ||p||=\sup_{k}|p_{k}|<\infty,\quad q_{k},p_{k}\in\mathbb{R}\}\label{banach_space}
\end{equation}
with the norm 
\[
||\psi||=\max\{||q||,||p||\}.
\]
Define also the bounded (non-selfadjoint) linear operator in $X$
\begin{equation}
A\psi=\psi',\quad\psi=(q,p),\ \psi'=(q',p'),\label{Adef}
\end{equation}
where 
\[
q'=p,\,\,\,p'_{k}=\omega^{2}(q_{k-1}-q_{k})-\alpha p_{k},\quad q_{0}=0
\]
Then the dynamics (\ref{x_main}), if $q_{k}=x_{k},p_{k}=\dot{x}_{k},k\geq1$
and $x_{0}(t)=0$, can be written as 
\begin{equation}
\dot{\psi}=A\psi\Longrightarrow\psi(t)=e^{At}\psi(0)\label{psi_dynamics}
\end{equation}
}

Proof of theorem \ref{Th_stability_4} and \ref{Th_stability_5} is
similar to the proof of theorems \ref{Th_stability_1} and \foreignlanguage{english}{\ref{Th_stability_3},}
but one should use, instead of lemma \ref{mainLemma}, the following
lemma.

\begin{lemma}Consider the system of equations 
\begin{equation}
\ddot{\frac{d^{2}q_{k}}{dt^{2}}+\alpha\frac{dq}{dt}+\omega^{2}q_{k}=\omega^{2}q_{k-1},k=1,2,...}\label{q_k_system}
\end{equation}
that is the same as (\ref{x_main}), where $q_{0}(t)$ is continuous
and absolutely integrable on $\mathbb{R}_{+}$: 
\[
\int_{0}^{+\infty}|q_{0}(t)|dt=Q<\infty.
\]
Assume the following initial conditions 
\[
\sum_{k=1}^{\infty}|q_{k}(0)|=a<\infty,\quad\sum_{k=1}^{\infty}|\dot{q}_{k}(0)|=b<\infty.
\]
If $\sqrt{2}\leqslant\alpha\leq2\omega$ then 
\[
\sup_{k,t}|q_{k}(t)|\leqslant Q'=2a+\frac{2b}{\omega}+2\omega Q.
\]
\end{lemma}

Proof. The system (\ref{q_k_system}) can be written in the operator
form, as in 
\begin{equation}
\dot{\psi}=A\psi+\omega^{2}q_{0}(t)g_{1},\ \psi(0)=\psi,\label{maineqOperator}
\end{equation}
where $g_{1}=(0,e_{1})^{T}\in X$, $e_{1}\in$ is the vector $(1,0,0,0...)$.
Then the solution of (\ref{maineqOperator}) is quite standard, see
\cite{D-K}, 
\[
\psi(t)=e^{tA}\psi(0)+\omega^{2}\int_{0}^{t}q_{0}(s)\ e^{(t-s)A}g_{1}\ ds.
\]
and 
\begin{equation}
e^{tA}=-\frac{1}{2\pi i}\int_{\Gamma}e^{zt}R(z)dz,\label{evolution_resolvent}
\end{equation}
where $\Gamma$ is a closed contour around the spectrum of $A$. Using
lemma \ref{Lemma_resolvent} we get 
\begin{equation}
q_{k}(t)=q_{k}^{(L)}(t)+q_{k}^{(N)}(t),\label{q_k_t_decompose}
\end{equation}
where 
\begin{align}
q_{k}^{(L)}(t)= & \ (e^{tA}\psi(0))(q_{k})=-\frac{1}{2\pi i}\sum_{j=0}^{k-1}\int_{\Gamma}\phi^{j}(z)l_{k-j}(z)e^{tz}dz,\label{qL}\\
q_{k}^{(N)}(t)= & \ \frac{\omega^{2}}{2\pi i}\int_{\Gamma}\frac{\phi^{k-1}(z)}{z^{2}+\alpha z+\omega^{2}}Q(t,z)dz,\quad Q(t,z)=\int_{0}^{t}e^{(t-s)z}q_{0}(s)ds.\label{qN}
\end{align}
Then change of variables gives 
\[
Q(t,z)=\int_{0}^{t}e^{sz}q_{0}(t-s)ds=\sum_{k=0}^{\infty}\frac{z^{k}}{k!}\int_{0}^{t}s^{k}q_{0}(t-s)ds.
\]
As for any $t\geqslant0$ 
\[
\left|\int_{0}^{t}s^{k}q_{0}(t-s)ds\right|\leqslant t^{k}Q,
\]
the $Q(t,z)$ is entire for any $t\geqslant0$. Note that the integrands
in the formulas (\ref{qL})-(\ref{qN}) are meromorphic functions
in the complex plane, having two exactly two poles $z_{1},z_{2}$,
corresponding to the multiplicity of the roots of the polynomial $G(z)$.
By theorem \ref{specTh}, the points $z_{1},z_{2}$ belong to the
spectrum of $A$, and by the same theorem, we can to choose the contour
$\Gamma$ so that the following 3 conditions hold
\begin{enumerate}
\item if $z\in\Gamma$ then $\mathrm{Re}(z)\leqslant0$;
\item $\sigma(A)\cap\Gamma=\{0\}$;
\item the points $z_{1},z_{2}$ lie inside the domain bounded by $\Gamma$.
\end{enumerate}
Using formulas (\ref{qL})-(\ref{qN}) for the chosen $\Gamma$ ,
we will get estimates of the functions $q_{k}^{(L)},q_{k}^{(N)}$.
It is clear that $|\phi(z)|\leqslant1$ for all $z\in\Gamma$. Then
\[
|q_{k}^{(L)}(t)|\leqslant\frac{1}{2\pi}\sum_{j=0}^{k-1}\int_{\Gamma}|\phi^{j}(z)||l_{k-j}(z)||e^{tz}||dz|\leqslant c'_{q}a+c'_{p}b,
\]
where the constants 
\[
c'_{q}=\frac{|\Gamma|}{2\pi}\frac{1}{\omega^{2}}\max_{z\in\Gamma}|\alpha+z|,\quad c'_{p}=\frac{|\Gamma|}{2\pi}\frac{1}{\omega^{2}}.
\]
Similarly we get estimate for $q_{k}^{(N)}$: 
\[
|q_{k}^{(N)}(t)|\leqslant Q\frac{|\Gamma|}{2\pi}.
\]
We want to estimate the length of the contour $\Gamma$, satisfying
the conditions 1-3. Assuming that $\alpha<2\omega$\c{ }let the points
$z_{1},z_{2}$ are enumerated so that 
\[
z_{1}=-\frac{\alpha}{2}+ir,\quad z_{2}=-\frac{\alpha}{2}-ir,\quad r=\sqrt{\omega^{2}-\frac{\alpha^{2}}{4}}.
\]
Then $z^{2}+\alpha z+\omega^{2}=(z-z_{1})(z-z_{2})$, and if $z\in\sigma(A)$,
then at least one of the inequalities $|z-z_{1}|\leqslant\omega$
or $|z-z_{2}|\leqslant\omega$ holds, see Theorm \ref{specTh} below.
It follows that $\sigma(A)$ belongs to the union of two circles 
\[
K_{1}=\{z\in\mathbb{C}:\ |z-z_{1}|\leqslant\omega\},\quad K_{2}=\{z\in\mathbb{C}:\ |z-z_{1}|\leqslant\omega\}.
\]
Note that both circles intersect the real axis at the points $-\alpha$
and $0$. That is why $\Gamma$ can be presented as the union of a
part of the circle bounding $K_{1}$, segment of the imaginary axis
and a part of the circle bounding $K_{2}$. Then 
\[
|\Gamma|\leqslant4\pi\omega,\quad\max_{z\in\Gamma}|z+\alpha|=2\omega.
\]
and finally 
\[
|q_{k}^{(L)}(t)|\leqslant2a+\frac{2b}{\omega},\quad|q_{k}^{(N)}(t)|\leqslant2\omega Q.
\]

The proof of Theorem \ref{Th_stability_5} is similar to the proof
of Theorem \ref{Th_stability_4}, only instead of the variables $q_{k}(t)$,
introduced therein, one should use variables $x_{k}(t)$, introduced
in (\ref{def_x_k}).

\selectlanguage{english}%

\subsection{Theorem \ref{th_dens}}

Rewriting $L_{N}$ in terms of $x_{k}$ 
\begin{equation}
L_{N}(t)=\frac{z_{0}(t)-z_{N}(t)}{N}=\frac{\sum_{k=1}^{N}x_{k}(t)}{N}+d,\,t\geqslant0\label{L_N_t}
\end{equation}
and using the equations (\ref{x_main}), we get the linear equation
for $L_{N}$\foreignlanguage{russian}{ 
\[
\ddot{L}_{N}+\alpha\dot{L}_{N}+\omega^{2}L_{N}=\frac{\omega^{2}}{N}\sum_{k=1}^{N}x_{k-1}+d\omega^{2}=\omega^{2}(\sum_{k=1}^{N}\frac{x_{k}}{N}+d+\frac{x_{0}-x_{N}}{N})=\omega^{2}L_{N}+\omega^{2}\frac{x_{0}-x_{N}}{N}
\]
or 
\begin{equation}
\ddot{L}_{N}+\alpha\dot{L}_{N}=f_{N}(t)=\omega^{2}\frac{x_{0}-x_{N}}{N}\label{eq_L_N_t}
\end{equation}
It is easy to see that the solution of this equation is 
\[
L(t)=L(0)+\frac{1}{\alpha}(1-e^{-\alpha t})\dot{L}(0)+g_{N}(t)
\]
where 
\[
g_{N}(t)=\int_{0}^{t}e^{-\alpha(t-s)}h_{N}(s)ds,\,\,\,h_{N}(s)=\int_{0}^{s}f_{N}(s')ds'
\]
\begin{lemma}\label{uniform_bound} There exists smooth function
$c(t)$ on $R_{+}$ such that for any $k\geq1$ and any $t\geq0$
\[
|x_{k}(t)|\leqslant c(t).
\]
\end{lemma}}

\selectlanguage{russian}%
Proof. The solution can be written as in (\ref{q_k_t_decompose})

\[
x_{k}(t)=q_{k}^{(L)}(t)+q_{k}^{(N)}(t).
\]

We can use formula (\ref{q_k_t_decompose}), choosing any contour
$\Gamma$, which does not intersect the spectrum of $A$. It is clear
that for any $z\in\Gamma$ we have $|\phi(z)|<q<1$ for some $q$.
Then 
\[
|q_{k}^{(L)}(t)|\leqslant\frac{|\Gamma|}{2\pi}\frac{\theta dc+\beta}{1-q}\frac{1}{\omega^{2}}u(t),\quad u(t)=\max_{z\in\Gamma}|e^{zt}|,\quad c=\max_{z\in\Gamma}|z+\alpha|.
\]
Similarly one can the bound for $q_{k}^{(N)}(t)$.

It follows from this lemma that for any $0<T<\infty$ the sequence
$f_{N}(t)\to0$, as $N\to\infty$, uniformly in $t\in[0,T]$. Thus
for any $t>0$ 
\[
\lim_{N\rightarrow\infty}g_{N}(t)=0
\]
and the theorem follows.

\section{Instability: proofs}

\selectlanguage{english}%

\subsection{Theorem \ref{th_instab_1}}

This is just an exercise to see that the reason for this is the resonance
effect: already $x_{1}(t)$ obtains harmonic components with frequencies
$\omega,\omega_{0}$, that implies resonance for the particle $2$
as its proper frequency is $\omega$. Here

\begin{equation}
\frac{d^{2}x_{1}(t)}{dt^{2}}+\omega^{2}x_{1}(t)=f(t)=-\frac{1}{(\omega_{0}\omega)^{2}}\frac{d^{2}z_{0}(t)}{dt^{2}}\label{x_1_eq}
\end{equation}
with $x_{1}(0)=\dot{x}_{1}(0)=0$. In case $\omega=\omega_{0}$ the
proper frequency coincides with the frequency of the external force,
and already the first particle will have resonance beha\t{v}iour that
is $\sup_{t\geq0}\,|x_{1}(t)|=\text{\ensuremath{\infty}}$. Consider
now the case when $\omega\neq\omega_{0}$. The equation (\ref{x_1_eq})
has the solution

\begin{equation}
x_{1}(t)=\frac{1}{\omega^{2}-\omega_{0}^{2}}(\sin(\omega_{0}t)-\frac{\omega_{0}}{\omega}\sin(\omega t))
\end{equation}
That is why already for the particle $2$ the resonance occurs. Note
however, that the first collision can occur between particles $k$
and $k-1$ for $k>2$.

\subsection{Spectrum, Theorem \ref{th_instab_2} and Corollary \ref{-Corollary_2}}

\paragraph{Remarks concerning spectrum}

\begin{theorem} \label{specTh} The spectrum $\sigma(A)$ of $A$
is the set 
\begin{equation}
\sigma(A)=\{z\in\mathbb{C}:\ |z^{2}+\alpha z+\omega^{2}|\leqslant\omega^{2}\}.\label{specFormula}
\end{equation}
As a corollary we have:

1) if $\alpha\geqslant\sqrt{2}\omega$ then for any $z\in\sigma(A)$
we have $\mathrm{Re}(z)\leqslant0$ and the equality takes place only
if $z=0$.

2) if $\alpha<\sqrt{2}\omega$ then $\sigma(A)$ contains some segment
of the imaginary axis together with a neighbourhood.\end{theorem}

Proof. Let\foreignlanguage{russian}{ 
\[
R(z)=(A-zE)^{-1},\ z\in\mathbb{C}
\]
be the resolvent of $A$. We use the following Lemma.}

\selectlanguage{russian}%
\begin{lemma}\label{Lemma_resolvent}\par Let $\psi=(q,p)\in X,\ \psi'=(q',p')\in X$
and 
\[
R(z)\psi=\psi'.
\]
Then for any $k=1,2,\ldots$ 
\[
q'_{k}=\sum_{j=0}^{k-1}\phi^{j}(z)l_{k-j},\,\,\,p'_{k}=q_{k}+zq'_{k}
\]
where 
\begin{equation}
\phi(z)=\frac{\omega^{2}}{z^{2}+\alpha z+\omega^{2}},\quad l_{k}(z)=-\frac{(\alpha+z)q_{k}+p_{k}}{z^{2}+\alpha z+\omega^{2}}\label{psi_of_z}
\end{equation}
\par \end{lemma}

Proof. From the definition 
\[
\psi=(A-zE)\psi'.
\]
Then $q=p'-zq'$ and $p'=q+zq'$. Moreover, we have the system of
equations 
\[
p_{k}=\omega^{2}(q'_{k-1}-q'_{k})-(\alpha+z)p'_{k}=\omega^{2}(q'_{k-1}-q'_{k})-(\alpha+z)(q_{k}+zq'_{k}).
\]
which, using (\ref{psi_of_z}), can be rewritten as 
\[
q'_{k}=\frac{\omega^{2}}{z^{2}+\alpha z+\omega^{2}}q'_{k-1}-\frac{(\alpha+z)q_{k}+p_{k}}{z^{2}+\alpha z+\omega^{2}}=\phi(z)q'_{k-1}+l_{k}(z),\quad.
\]
Finally 
\begin{equation}
q'_{k}=\sum_{j=0}^{k-1}\phi^{j}(z)l_{k-j}\label{resEq}
\end{equation}

Proof of Theorem \ref{specTh}.

Let us consider two cases. Let $|\phi(z)|=q<1$. Note that for any
$k$ 
\[
|l_{k}(z)|\leqslant|\frac{(\alpha+z)+1}{z^{2}+\alpha z+\omega^{2}}|||\psi||=c.
\]
It follows that for any $k$ 
\[
|q'_{k}|\leqslant c\frac{1-q^{k}}{1-q}\leqslant\frac{c}{1-q}.
\]
Then the operator $R(z)$ is bounded and $z\notin\sigma(A)$. Another
possibility is $|\phi(z)|>1$. Consider the sequence $\psi_{n}=(q^{n},p^{n})\in X$,
where 
\begin{align*}
q^{n}= & \ 0\\
p_{k}^{n}= & \ \begin{cases}
1, & k<n\\
0, & k\geqslant n.
\end{cases}
\end{align*}
Denote $\psi'_{n}=R(z)\psi_{n},\ \psi'_{n}=(\hat{q}^{n},\hat{p}^{n})$.
From formula (\ref{resEq}) we get 
\[
\hat{q}_{n}^{n}=-\frac{1}{z^{2}+\alpha z+\omega^{2}}\frac{1-\phi^{n}(z)}{1-\phi(z)}
\]
That is why for $|\phi(z)|>1$ 
\[
||R(z)\psi_{n}||\geqslant\frac{1}{|z^{2}+\alpha z+\omega^{2}|}\left|\frac{1-\phi^{n}(z)}{1-\phi(z)}\right|\rightarrow\infty,\ \mbox{\textcyr{\char239}\textcyr{\char240}\textcyr{\char232}}\ n\rightarrow\infty.
\]
However, it is evident that $||\psi_{n}||=1$ and thus $R(z)$ is
not bounded. We conclude then that $z\in\sigma(A)$. As the spectrum
is a closed subset, we get the proof of the first part of the theorem.
Now let us prove the two other assertions.

For $z=a+ib,\ a,b\in\mathbb{R}$ we have 
\[
z^{2}+\alpha z+\omega^{2}=a^{2}-b^{2}+\alpha a+\omega^{2}+i(2ab+\alpha b).
\]
Then the inequality $|z^{2}+\alpha z+\omega^{2}|\leqslant\omega^{2}$
is equivalent to the inequality 
\[
(a^{2}+\alpha a+\omega^{2}-b^{2})^{2}+b^{2}(2a+\alpha)^{2}\leqslant\omega^{4}.
\]
or 
\begin{equation}
b^{4}+b^{2}((2a+\alpha)^{2}-2(a^{2}+\alpha a+\omega^{2}))+(a^{2}+\alpha a+\omega^{2})^{2}-\omega^{4}\leqslant0.\label{specineq}
\end{equation}
If we denote 
\begin{align*}
A= & \ (2a+\alpha)^{2}-2(a^{2}+\alpha a+\omega^{2})=2a^{2}+2\alpha a+\alpha^{2}-2\omega^{2},\\
B= & \ (a^{2}+\alpha a+\omega^{2})^{2}-\omega^{4}=a(a+\alpha)(a^{2}+\alpha a+2\omega^{2}).
\end{align*}
the inequality (\ref{specineq}) can be rewritten as 
\[
h(a,b)=b^{4}+Ab^{2}+B\leqslant0.
\]
Consider the case $\alpha\geqslant\sqrt{2}\omega$. Then for any $a>0$
and $b\in\mathbb{R}$ we have evidently $h(a,b)>0$, that is $z\notin\sigma(A)$.
If $a=0$ then $B=0$ and $A\geqslant0$. It follows that the unique
$b$ such that $h(0,b)\leqslant0$ is $b=0$. Thus the first assertion
is proved.

Let now $\alpha<\sqrt{2}\omega$. For $a=0$ we have $A=\alpha^{2}-2\omega^{2}<0,\ B=0$.
It follows that there exists $b\in[-\sqrt{-A},\sqrt{-A}]$ such that
$h(0,b)\leqslant0$. From continuity of $h$ it follows that there
is some interval of the imaginary axis around this $b$, belonging
to the spectrum.

\selectlanguage{english}%
\begin{remark} Stability problems, considered here, cannot be completely
treated by spectral methods. In fact, the stability concern the value
of $\sup_{t\geqslant0}||q(t)||$. This question was studied by many
authors, in particular, by Daletsky and Krein, see (\cite{D-K}).
Their results concern the case of $\epsilon$-dichotomic operators,
i.e. when the spectrum of the operator $A$ is the union of two subsets
$\sigma_{+},\sigma_{-}$ of the complex plane, such that $\sigma_{+}$
belongs to the open right half-plane and $\sigma_{-}$ belonds to
the open left half\={p}lane. We have shown that for any parameters
the spectrum contain the point $0$ and thus $A$ is not $\epsilon$-dichotomic
operator.

\end{remark}

\paragraph{Proof of Theorem \ref{th_instab_2}}

\selectlanguage{russian}%
Using formula (\ref{evolution_resolvent}), lemma \ref{Lemma_resolvent}
and the theorem assumptions one gets: 
\[
q_{k+1}(t)=-\frac{\epsilon}{2\pi i}\int_{\Gamma}e^{zt}\phi^{k}(z)dz.
\]
The integrand has two poles (roots of the characteristic equation).
Because of $\alpha<\sqrt{2}\omega$ the roots are: 
\[
\lambda_{1}=-\frac{\alpha}{2}+ir,\ \lambda_{2}=-\frac{\alpha}{2}-ir,\quad r=\sqrt{\omega^{2}-\frac{\alpha^{2}}{4}}
\]
By Jordan lemma, for any $a>-\frac{\alpha}{2}$ 
\[
q_{k+1}(t)=-\frac{\epsilon}{2\pi i}\int_{a-i\infty}^{a+i\infty}e^{zt}\phi^{k}(z)dz.
\]
For $t=\mu k$ we can rewrite this as 
\begin{align}
 & q_{k+1}(\mu k)=F(\mu,k)=-\frac{\epsilon}{2\pi i}\int_{a-i\infty}^{a+i\infty}\exp\left(k(\mu z-\ln(z^{2}+\alpha z+\omega^{2})+\ln\omega^{2})\right)dz=\\
 & =-\frac{\epsilon}{2\pi i}\int_{a-i\infty}^{a+i\infty}\exp(kS(z))dz,\quad S(z)=\mu z-\ln(z^{2}+\alpha z+\omega^{2})+\ln\omega^{2}.\label{integX}
\end{align}

The mapping $z^{2}+\alpha z+\omega^{2}$ transforms the line $a+ib,\ b\in\mathbb{R}$
to parabola. Thus one can select holomorphic branch of the logarithm
on any such line with $a>-\frac{\alpha}{2}$ 
\[
\ln(z^{2}+\alpha z+\omega^{2})=\ln|z^{2}+\alpha z+\omega^{2}|+i\arg(z^{2}+\alpha z+\omega^{2}).
\]
Now we use the saddle point method. Saddle points are found out from
the equation: 
\[
S'(z)=\mu-\frac{2z+\alpha}{z^{2}+\alpha z+\omega^{2}}=0.
\]
or from the quadratic equation 
\begin{equation}
\mu z^{2}+(\alpha\mu-2)z+\omega^{2}\mu-\alpha=0.\label{chareq}
\end{equation}
The discriminant 
\[
D=(\alpha\mu-2)^{2}-4\mu(\omega^{2}\mu-\alpha)=\alpha^{2}\mu^{2}+4-4\mu^{2}\omega^{2}=-4\mu^{2}\tau^{2}+4.
\]
and we assume that 
\[
\mu>\frac{1}{\tau}.
\]
Then $D<0$ and there are two complex saddle points: 
\[
z_{\pm}=z_{\pm}(\mu)=\frac{2-\alpha\mu\pm i\sqrt{-D}}{2\mu}=-\frac{\alpha}{2}+\frac{1}{\mu}\pm\frac{i}{\mu}\sqrt{\mu^{2}\tau^{2}-1}.
\]
To check that $z_{\pm}(\mu)$ are simple saddle points, we should
find the second derivative 
\begin{align*}
 & S''(z_{\pm}(\mu))=-\frac{2}{z^{2}+\alpha z+\omega^{2}}+\frac{(2z+\alpha)^{2}}{(z^{2}+\alpha z+\omega^{2})^{2}}|_{z=z_{\pm}(\mu)}=-\frac{2\mu}{2z_{\pm}+\alpha}+\mu^{2}=\\
 & =-\frac{\mu}{\frac{1}{\mu}\pm\frac{i}{\mu}\sqrt{\mu^{2}\tau^{2}-1}}+\mu^{2}=\mu^{2}\left(\frac{\pm i\sqrt{\mu^{2}\tau^{2}-1}}{1\pm i\sqrt{\mu^{2}\tau^{2}-1}}\right).
\end{align*}
This shows that $S''(z_{\pm}(\mu))\ne0$. We shall put $a=a(\mu)=-\frac{\alpha}{2}+\frac{1}{\mu}$
in the formula (\ref{integX}) for $F(\mu,k)$ 
\[
F(\mu,k)=-\frac{\epsilon}{2\pi i}\int_{a(\mu)-i\infty}^{a(\mu)+i\infty}\exp(kS(z))dz.
\]
To use the line $a(\mu)+ib,\ b\in\mathbb{R}$ as the contour for the
saddle point method one should check two following conditions 
\begin{enumerate}
\item The function 
\[
H(y)=\mathrm{Re}(S(a+iy))
\]
reaches its maximum only at two points: $\mathrm{Im}(z_{\pm})=\pm\frac{1}{\mu}\sqrt{\mu^{2}\tau^{2}-1}.$
In fact 
\begin{align*}
 & H(y)=\mu a-\frac{1}{2}\ln\left((a^{2}-y^{2}+\alpha a+\omega^{2})^{2}+((2a+\alpha)y)^{2}\right)+\ln\omega^{2}=\\
 & =\mu a+\ln\omega^{2}-\frac{1}{2}\ln h(y^{2}),\quad h(s)=(s-(a^{2}+\alpha a+\omega^{2}))^{2}+(2a+\alpha)^{2}s.
\end{align*}
And it is sufficient to check that the graph of $H(y)$ is the parabola
with the vertex: 
\begin{align*}
 & s_{0}=-\frac{(2a+\alpha)^{2}-2(a^{2}+\alpha a+\omega^{2})}{2}=-\frac{\frac{4}{\mu^{2}}-2(\frac{\alpha^{2}}{4}-\frac{\alpha}{\mu}+\frac{1}{\mu^{2}}-\frac{\alpha^{2}}{2}+\frac{\alpha}{\mu}+\omega^{2})}{2}=\\
 & =-\frac{\frac{2}{\mu^{2}}+\frac{\alpha^{2}}{2}-2\omega^{2}}{2}=\tau^{2}-\frac{1}{\mu^{2}}=\mathrm{Im}(z_{\pm})^{2}
\end{align*}

\item In some neighbourhoods of the saddle point $z_{\pm}$ the contour
goes through two different sectors where $\mathrm{Re}(S(z))<\mathrm{Re}(S(z_{\pm}))$.
To prove this put 
\[
\nu=\sqrt{\mu^{2}\tau^{2}-1}
\]
and rewrite the expression for $S''(z_{\pm})$: 
\begin{equation}
S''(z_{\pm})=\mu^{2}\frac{\pm i\nu}{1\pm i\nu}=\mu^{2}\frac{\nu^{2}\pm i\nu}{1+\nu^{2}}=\frac{\nu}{\tau^{2}}(\nu\pm i).\label{eqSpp}
\end{equation}
Further on we agree that the arguments of complex numebrs take values
in the interval $(-\pi,\pi]$. Using (\ref{eqSpp}) we obtain: 
\[
\mathrm{arg}(S''(z_{+}))\in(0,\frac{\pi}{2}),\quad\mathrm{arg}(S''(z_{-}))\in(-\frac{\pi}{2},0).
\]
The argument of the axis of the saddle point (p. 84 \cite{Bruijn})
$z_{\pm}$ is 
\[
\phi_{\pm}=\frac{\pi}{2}-\frac{1}{2}\mathrm{arg}(S''(z_{\pm})).
\]
It follows that 
\[
\phi_{+}\in(\frac{\pi}{4},\frac{\pi}{2}),\quad\phi_{-}\in(\frac{\pi}{2},\frac{3\pi}{4}).
\]

\end{enumerate}
In both cases the angle between the integration contour (i.e. the
line $a(\mu)+ib$) and the saddle point axis is less then $\frac{\pi}{4}$.
That means that the line $a(\mu)+ib$ can be used as the saddle point
method contour (see for example,\c{ }p. 89 of\cite{Bruijn}).

So by formula $(5.7.2)$, p. $88$ of \cite{Bruijn}, we have the
asymptotic formula 
\begin{equation}
F(\mu,k)\sim-\frac{\epsilon}{2\pi i}\left(a_{+}e^{kS(z_{+})}+a_{-}e^{kS(z_{-})}\right),\ \mathrm{as}\ k\rightarrow\infty,\label{asympF}
\end{equation}
where the constants 
\[
a_{\pm}=\sqrt{\frac{2\pi}{k|S''(z_{\pm})|}}s_{\pm},\quad s_{\pm}=\exp{i\phi_{\pm}}.
\]
We choose the signs of the constants $s_{\pm}$ so that the angle
between the saddle point axis (this angle is determined with an accuracy
to $\pi$) and the contour of integration were acute (see \cite{Bruijn}).

One can rewrite the right part of the formula (\ref{asympF}) 
\[
S(z_{\pm})=\mu z_{\pm}-\ln\left(\frac{2z_{\pm}+\alpha}{\mu}\right)+\ln\omega^{2}=\mu z_{\pm}-\ln\left(\frac{2}{\mu^{2}}(1\pm i\sqrt{\mu^{2}\tau^{2}-1})\right)+\ln\omega^{2}
\]
Obviously, 
\[
\exp(kS(z_{-}))=\overline{\exp(kS(z_{+}))}.
\]
and $\mathrm{arg}(S''(z_{-}))=-\mathrm{arg}(S''(z_{+}))$. Then 
\[
\overline{s_{-}}=\exp(-i\phi_{-})=\exp(-i(\frac{\pi}{2}+\mathrm{arg}(S''(z_{+})))=\exp(-i\pi)\exp(i\phi_{+})=-s_{+}.
\]
Here using (\ref{eqSpp}) we get 
\[
|S''(z_{+})|=|S''(z_{-})|=\frac{\nu}{\tau^{2}}\sqrt{\nu^{2}+1}=\frac{\nu\mu}{\tau}.
\]
and 
\begin{equation}
F(\mu,k)\sim-\frac{\epsilon}{2\pi i}\left(a_{+}\exp(kS(z_{+}))-\overline{a_{+}\exp(kS(z_{+}))}\right)=-\frac{\epsilon}{\pi}\mathrm{Im}(a_{+}\exp(kS(z_{+}))\label{fmukim}
\end{equation}

\[
\mathrm{Re}(S(z_{+}))=\mu a(\mu)-\ln\left|\frac{2}{\mu^{2}}(1+i\sqrt{\mu^{2}\tau^{2}-1})\right|+\ln\omega^{2}=\mu a(\mu)-\ln\left(\frac{2\tau}{\mu}\right)+\ln\omega^{2}=f(\mu)
\]
\[
\mathrm{arg}\left(1+i\sqrt{\mu^{2}\tau^{2}-1}\right)=\arg S^{''}(z_{-})+\frac{\pi}{2}=\phi_{+}
\]
\[
\mathrm{Im}(S(z_{+}))=\sqrt{\mu^{2}\tau^{2}-1}-\phi_{+}=\nu-\phi_{+}
\]
Substituting this to formula (\ref{fmukim}), we get 
\[
F(\mu,k)\sim-\epsilon\sqrt{\frac{2\tau}{\pi k\nu\mu}}e^{kf(\mu)}\mathrm{Im}(e^{i(\phi_{+}+k(\nu-\phi_{+}))})=\epsilon\sqrt{\frac{2\tau}{\pi k\nu\mu}}e^{kf(\mu)}\sin(\Omega(\mu)k+\phi_{0}(\mu))
\]
where 
\[
\phi_{0}(\mu)=\phi_{+}=\arctan(\nu),\,\,\,\Omega(\mu)=\nu-\phi_{+}=\nu-\arctan(\nu)
\]

\selectlanguage{english}%

\paragraph{Corollary \ref{-Corollary_2}}

We need the following 

\begin{lemma}\label{f_mu}\foreignlanguage{russian}{ If $\alpha\leq\sqrt{2}\omega$
there exists $\delta>0$ such that for all $\mu\in(\frac{1}{r},\frac{1}{r}+\delta)$
\[
f(\mu)>0.
\]
}

\selectlanguage{russian}%
\end{lemma}

Proof. Using $\tau=\sqrt{\omega^{2}-\frac{\alpha^{2}}{4}}$ and $\alpha<\sqrt{2}\omega$
we get 
\[
f\left(\frac{1}{\tau}\right)=-\frac{\alpha}{2\tau}-\ln\left(\frac{\tau^{2}}{\omega^{2}}\right)+1-\ln2=1-\ln2-\frac{x}{\sqrt{1-x^{2}}}-\ln(1-x^{2})=h(x),\ x=\frac{\alpha}{2\omega}<\frac{1}{\sqrt{2}}
\]
where obviously 
\[
h(0)=1-\ln2>0,\quad h\left(\frac{1}{\sqrt{2}}\right)=0.
\]
The following calculation 
\begin{align*}
 & h'(x)=-\frac{1}{1-x^{2}}-\frac{x^{2}}{(1-x^{2})\sqrt{1-x^{2}}}+\frac{2x}{1-x^{2}}=\frac{-\sqrt{1-x^{2}}-x^{2}+2x\sqrt{1-x^{2}}}{(1-x^{2})\sqrt{1-x^{2}}}=\\
 & =-\frac{(x-\sqrt{1-x^{2}})^{2}}{(1-x^{2})\sqrt{1-x^{2}}}<0,\ \mbox{for}\ x<1.
\end{align*}
shows that the function $h$ is monotone decreasing on $[0,1)$. As
the function $f$ is smooth, the lemma is proved.

For $t=\mu k,\ k\rightarrow\infty$ we have: 
\begin{align*}
 & R_{k}=r_{k+1}(\mu k)=z_{k}(t)-z_{k+1}(t)=a+q_{k}-q_{k+1}\sim\\
 & \sim a+\frac{c}{\sqrt{k-1}}e^{(k-1)f(\mu)}\left(\sin(\Omega(\mu)(k-1)+\phi_{0})-\sqrt{\frac{k-1}{k}}e^{f(\mu)}\sin(\Omega(\mu)k+\phi_{0})\right)
\end{align*}
We chose $\mu$ so that $f(\mu)>0$ and $\Omega(\mu)$ is irrational.
There exists a subsequence $k_{n},n=1,2...,$ such that $k_{n}\rightarrow\infty$
as $n\rightarrow\infty$ and 
\[
\liminf_{n\rightarrow\infty}\sin(\Omega k_{n}+\phi_{0})=-1,\quad\limsup_{n\rightarrow\infty}\sin(\Omega k_{n}+\phi_{0})=1
\]
Obviously then: 
\[
\liminf_{n\rightarrow\infty}R_{k_{n}}=-\infty,\quad\limsup_{n\rightarrow\infty}R_{k_{n}}=+\infty
\]
Corollary \foreignlanguage{english}{\ref{-Corollary_2}} is proved. 
\selectlanguage{english}%

\end{document}